\documentclass[]{spie}  

\usepackage{amsmath,amsfonts,amssymb}
\usepackage{graphicx}
\usepackage[colorlinks=true, allcolors=blue]{hyperref}
\usepackage{comment}

\title{Advancements in Streamlining Time-Domain and Multi-Messenger Astronomy Follow-Up Infrastructure at Keck Observatory}

\author[a]{Chien-Hsiu Lee}
\author[a]{Jeffrey A. Mader}
\author[a]{Tyler Coda}
\author[a]{Jo Hayashi}
\author[a]{Max Brodheim}
\author[a]{Lucas Fuhrman}

\affil[a]{W. M. Keck Observatory, 65-1120 Mamalahoa Hwy, Kamuela, HI 96720, USA}

\authorinfo{Further author information: (Send correspondence to C.-H. Lee, currently based at Hobby-Eberly telescope)\\C.-H. Lee: E-mail: chienhsiu.lee@austin.utexas.edu}

\begin{document} 
\maketitle

\begin{abstract}
With active time-domain surveys like the Zwicky Transient Facility, the anticipated Rubin Observatory's Legacy Survey of Space and Time, and multi-messenger experiments such as LIGO/VIRGO/KANGRA for gravitational wave detection and IceCube for high-energy neutrino events, there is a new era in both time-domain and multi-messenger astronomy. The Astro2020 decadal survey highlights effectively responding to these astronomical alerts in a timely manner as a priority, and thus, there is an urgent need for the development of a seamless follow-up infrastructure at existing facilities that are capable of following up on detections at the surveys’ depths. At the W. M. Keck Observatory (WMKO), we are actively constructing critical infrastructure, aimed at facilitating the Target-of-Opportunity (ToO) trigger, optimizing observational planning, streamlining data acquisition, and enhancing data product accessibility. In this document, we provide an overview of these developing services and place them in context of existing observatory infrastructure like the Keck Observatory Archive (KOA) and Data Services Initiative (DSI).
\end{abstract}

\keywords{Time-domain astronomy, astronomy software, computational methods}

\section{INTRODUCTION}
\label{sec:intro}  

Large format digital cameras, coupled with wide-field telescopes, have ushered in a new era of time-domain astronomy with high-cadence all-sky surveys, e.g. Catalina Real-Time Transient Survey\cite{2009ApJ...696..870D}, Palomar Transient Factory\cite{2009PASP..121.1395L}, All-Sky Automated Survey for SuperNovae\cite{2014ApJ...788...48S}, Pan-STARRS\cite{2016arXiv161205560C}, Asteroid Terrestrial-impact Last Alert System\cite{2018PASP..130f4505T}. Among these surveys, the most notable project is the Zwicky Transient Facility \cite{2019PASP..131a8002B} (ZTF), successor of the Palomar Transient Factory, which carries out a public survey funded by the National Science Foundation (NSF) to patrol the northern sky every two to three nights in the Sloan g- and r-band filters, covering 47 deg$^2$ in a single shot with the wide-field camera mounted on the Palomar 48-inch telescope since 2018. The high etendue of ZTF results in more than 10$^5$ alerts on a nightly basis with objects as faint as r = 20.5 mag. 

The large number of alerts generated each night by the aforementioned surveys have revealed numerous intriguing astronomical objects. On the transient end, ZTF has pushed the boundary of fast transients\cite{2020ApJ...895...49H,2020arXiv200610761H}, calcium-rich transients\cite{2020arXiv200409029D}, as well as tidal disruption events\cite{2020arXiv200101409V}. 
The high-cadence, long-term monitoring photometric data from ZTF also benefit studies of variable stars, including but not limited to eruptive variables\cite{2020arXiv201014679S}, long-term variables from e.g. R Coronae Borealis stars\cite{2020AJ....159...61L}, as well as ultra-short period eclipsing binaries composed of double white dwarfs\cite{2019Natur.571..528B} as potential sources of gravitational wave experiments like the Laser Interferometer Space Antenna (LISA). 
Data from the high etendue of ZTF also flourishes discoveries of Solar System objects, e.g. the first Vatira object\cite{2020arXiv200904125I} -- asteroids with orbits within that of Venus -- as well as extensive studies of the first interstellar comet 2I/Borisov \cite{2020AJ....159...77Y}. 
Wide-field optical surveys like ZTF play crucial roles in multi-messenger astronomy -- especially when the localization of gravitational wave events is poorly defined -- to identify optical counterparts of gravitational wave events before they fade away\cite{2020NatRP...2..452K}. ZTF also jointly discovered tidal disruption events that coincided with high energy neutrinos\cite{2020arXiv200505340S}.
When the Vera C. Rubin Observatory\cite{2019ApJ...873..111I} comes online, its Legacy Survey of Space and Time will provide an unprecedented time-domain survey with its wide-field (9.6 deg$^2$ field-of-view) camera on the 8.4 meter Simonyi Survey Telescope on the  El Pe\~{n}\'on summit of Cerro Pach\'on. Particularly, the Rubin Observatory is expected to deliver 10 million alerts per night, uncovering events as faint as $\sim$24 mag, beyond the reach of ZTF.  

These examples provide a flavor of the exciting and cutting-edge science that can be done with ongoing and future time-domain surveys. With the flood of alerts generated by these surveys, it is not feasible for astronomers to visually inspect each individual alert. Instead, the time-domain community came up with the idea of astronomical event brokers, similar to the idea of broker in exchange, to sift through these time-sensitive events and select the most interesting objects that warrant follow-up. The alert brokers are critical as the follow-up resources, especially spectroscopic instruments on large telescopes, are unmatched to the number of alerts. 

In order to tackle  the flood of millions of alerts from ZTF (and Rubin Observatory in the future), it is imperative to have a software infrastructure that can sift through alerts. There are several community alert brokers, such as the the Arizona-NOIRLab Temporal Analysis and Response to Events System \cite{2014SPIE.9149E..08S,2016SPIE.9910E..0FS,2018ApJS..236....9N} (ANTARES) developed by the University of Arizona and NSF's National Optical-Infrared Astronomy Research Laboratory (NOIRLab), ALeRCE by the Chilean community, AMPEL funded by the German government, FINK by the French community, and LASAIR by the UK consortium.

Once flagged by the brokers, the next step is to coordinate follow-up observations with the limited telescope resources in a timely manner. This can be done by sharing target and observation resources in a common platform, such as the Target and Observation Manager (TOM) toolkit developed by the Las Cumbres Observatory \cite{2018SPIE10707E..11S} or the SkyPortal astronomical data platform \cite{skyportal2019}.

The benefit of using these common platforms is they allow users to specify their access to different telescope resources and data archives, to provide a one-stop service to trigger observations and gather and share data among collaborators, instead of contacting each individual telescope and observatory archive. The TOMtoolkit is an integral part of the Astronomical Event Observatory Network \cite{2020SPIE11449E..25S} (AEON), a protocol for Rubin Observatory in-kind telescope access. SkyPortal is heavily used by ZTF collaborations for dedicated follow-up. 

To actively engage in the time-domain community, Keck is developing and maturing software infrastructure that can connect to these common platforms. This includes a target-of-opportunity request tool and API for ToO triggering, as well as a python-based client to access data holdings at the Keck Observatory Archive (KOA). These services are detailed in the following sections.

\section{Infrastructure at Keck Observatory}
\subsection{Mode of operation}
Atop Maunakea in Hawaii, the W. M. Keck Observatory has patrolled the northern sky since 1993 with the inception of the Keck I telescope, and the addition of the Keck II telescope in 1996. Both telescopes have primary mirrors composed of 36 hexagonal segments with an equivalent size of 10 meter diameter. Equipped with a suite of cutting-edge optical and infrared instruments, Keck is well poised to respond to time sensitive astronomical events, such as transients, moving objects, and multi-messenger events. To support these increasing science observations, Keck installed a deployable tertiary mirror (K1DM3\cite{2018SPIE10706E..11R}) on the Keck I telescope in 2018 that allows instrument changes to happen more easily during the night.  Prior to the upgrade, only the installed Cassegrain instrument or, if the tertiary is installed, the two tertiary instruments were available on a given night.  The reconfiguration of instruments during the day also required significant resources.  With K1DM3 the number of instrument reconfigurations decreased and up to four Keck I instruments are available every night.  Keck II has up to three instruments available if its tertiary is installed.

Unlike other large telescopes, such as Gemini, GTC, and VLT, Keck operates in a classically scheduled mode. Twice yearly, Keck principal investigators (PIs) submit their proposals for review by the appropriate Telescope Allocation Committee (TAC).  If awarded observation time, the PI then plans and executes their observations.  PIs of classical modes have utilized Keck and made significant contributions to various facets of TDA/MMA, such as supernova cosmology, gamma-ray bursts, tidal disruption events, gravitational wave optical counterpart, and counterpart of high energy neutrino events.

\subsection{Target-of-Opportunity Observation Request Tool}
 For ToO observations, Keck PIs at similar institutions used to form collaborations and share observation time on pre-scheduled nights. With the addition of K1DM3 on Keck I, the Keck Observatory board approved a ToO policy that allows PIs more flexibility to submit ToO observations and interrupt other PIs' scheduled nights when consented to. There is also a limit of ToO triggers within the same partner institute or cross-partner, to properly compensate time among partner institutions. In light of this policy, Keck implemented the Target-of-Opportunity Observation Request Tool (TORT), a web form that facilitates interruption of the normal observing schedule for special ToO observations.  The TORT allows the PI, or any of their CO-Investigators, with allocated ToO hours to submit a request to perform the required observation.  The submission will initiate the process, shown in Figure 1, by which the observing time is scheduled, email notifications are sent out, and the progress of the interrupt is logged.  A ToO interrupt can be triggered at any time of the day or night.
 
 While the use of the TORT has been very successful, it still requires a PI to manually enter program information, target coordinates and properties, and instrument configurations, which will slow down the trigger process with the hundred thousands or millions of alerts per night delivered by the current and future surveys. To allow automatic trigger and reduce human interaction, we have implemented the TORT API that allows a PI to submit their ToO requests programmatically and promptly without human interaction.

\begin{figure}
\centering
\includegraphics[width=0.9\textwidth]{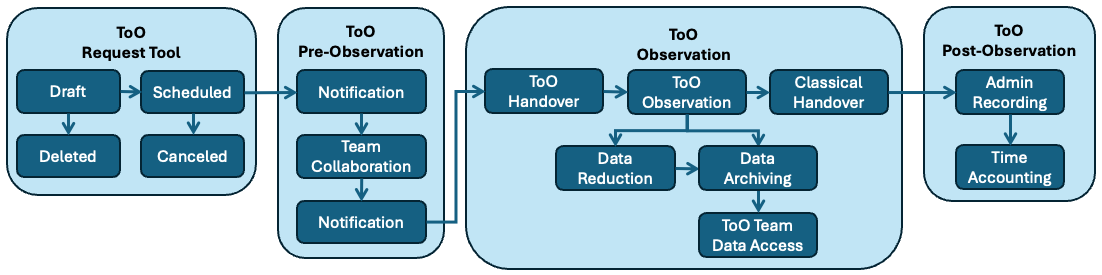}
\caption{Flow chart showing sequence of events from ToO request to observation.}
\label{fig:web}
\end{figure}

\subsection{Keck Observatory Archive}
Once an observation is finished, the raw data will be transferred from the summit to KOA at the Infrared Processing and Analysis Center (IPAC) in California. The requirement is for data to be transferred within 5 minutes after the science file is written to disk, however, most of the raw data are transferred within 60 seconds or less except for instruments with large format CCD. Besides the raw data, Keck also reduces the data with well-established data reduction pipelines (DRPs), most of them contributed by the community or by the instrument building team. A list of data reduction software, and the ones adopted by Keck, can be found at the Keck Data Reduction Pipeline page\footnote{https://www2.keck.hawaii.edu/inst/drp.html}. If DRPs are incorporated and proper calibration data are taken, Keck will also deliver quick-look (requirement within 5 minutes) and science-ready (requirement next day) data to KOA. These higher level data products are also searchable at KOA. 

Users can interact with KOA via three different avenues:
\begin{itemize}
    \item \textit{KOA website}\\
    The KOA landing page\footnote{https://koa.ipac.caltech.edu} allows users to search and browse public and proprietary data. Users can, for example, search for observations at a given position within a given search radius. Furthermore, users can specify the timing of the observations, or an extended period of time. Users can select different instruments and modes (imaging or spectroscopic). The KOA website also allows users, specifically PIs, to grant proprietary data access to collaborators. 
    \item \textit{Observers Data Access Portal}\\ 
    When carrying out observations on-site or remotely, users have the option to stream data directly to their computer in real-time during the night. This can be done via the Observers' Data Access Portal (ODAP), which was released in 2023.  ODAP will associate the PI and observers of that night with the corresponding instruments and scheduled programs. Users have the option to turn on real-time downloading for the different levels of data products as they are produced. Data will be stored in the default download directory specified by their web browser. 
    
    \item \textit{PyKOA}\\
    Another route to connect to KOA is via a Python-based client PyKOA\footnote{https://koa.ipac.caltech.edu/UserGuide/PyKOA/PyKOA.html}. Users can use PyKOA to query and download public and proprietary data. In addition, the client enables users to query the KOA database in a programmatic manner, compatible with VO and ADQL syntax. With access to the KOA database, users can also use the PyKOA client to develop customized notebooks to query and explore data in an interactive manner. 
    
\end{itemize}


\subsection{Connecting to downstream services}

\begin{figure}
\centering
\includegraphics[width=0.9\textwidth]{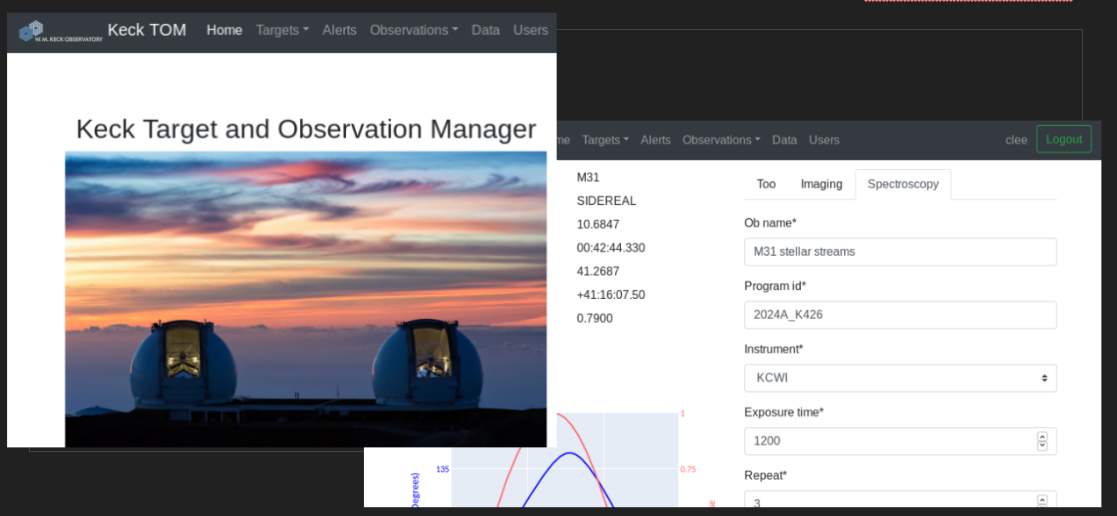}

\caption{Keck TOM developed with TOMtoolkit}
\label{fig:kecktom}
\end{figure}

In the current and future time-domain and multi-messenger ecosystem, observatories are not only receiving triggers from the surveys and alert brokers, but also need to provide an interface to science platforms for observers to work collaboratively. In this regard, Keck is developing infrastructure to connect to a suite of services that are already implemented by the astronomical community.
At Keck, we are implementing a Keck Target and Observation Manager (TOM), which is based on the TOMtoolkit\cite{2018SPIE10707E..11S} developed by Las Cumbres Observatory. Users will be able to use the Keck TOM to plan and trigger observations, retrieve observed data, and share it with collaborators, as shown in Figure \ref{fig:kecktom} Via TOMtoolkit, Keck is compatible to the Astronomical Event Observatory Network\cite{Street2020} (AEON)\footnote{http://ast.noao.edu/data/aeon}, a network of facilities including (but not limited to) the Las Cumbres Observatory, the Southern Astrophysical Research (SOAR) Telescope, and the Gemini Observatory that can provide coordinated observations. AEON is the preferred model for Rubin Observatory in-kind telescope access. 

Besides TOMtoolkit, Keck will also be connected to SkyPortal, an astronomical data platform\footnote{https://skyportal.io} heavily used by the ZTF collaboration for follow-ups.

\subsection{Science examples}
\label{sec:sci}

\begin{figure}
\centering
\includegraphics[width=0.9\textwidth]{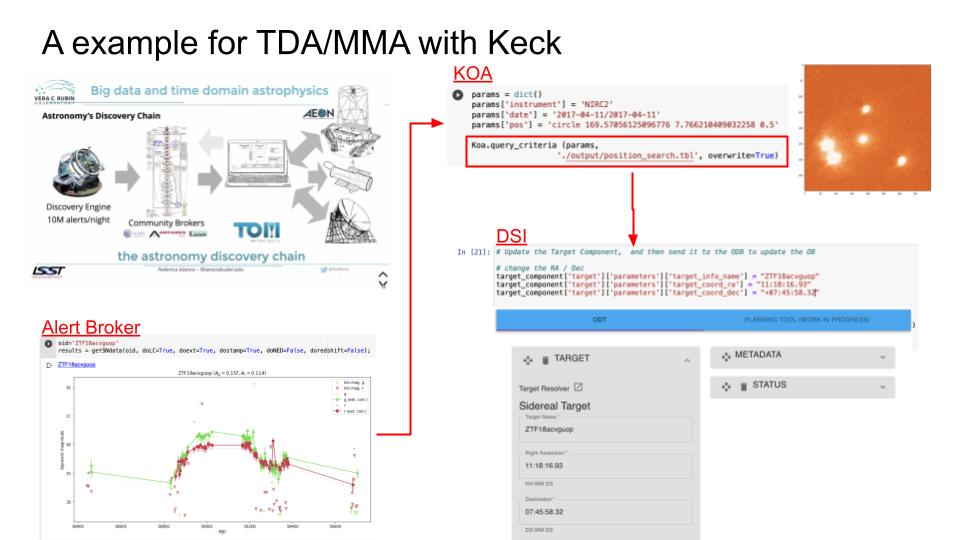}

\caption{Flow chart from surveys to brokers and triggering with Keck}
\label{fig:too}
\end{figure}

To illustrate how the time-domain and multi-messenger community can use Keck for time-sensitive follow-up, we show a workflow in Figure \ref{fig:too}. In this particular case, we use Rubin observatory as the upstream survey and alert provider. The alerts are then digested by the community alert brokers, which all have implemented python clients for users to interact. After an intriguing alert is identified, users can use PyKOA to query Keck archival data. Should the alert warrant follow-up, users can trigger a ToO observation with the TORT tool under Keck's Data Services Initiative (DSI) framework. 

Nowadays, surveys are providing hundred thousands to millions of alerts on a nightly basis, beyond the capability for human inspection to digest. Hence there are community brokers that will sift through these events to select the most interesting objects. In the first few years of Rubin observatory, before completing the planned 10-year survey, the image may not be deep enough to reveal the host galaxies of transients. Keck, as one of the largest telescopes in the world, has accumulated decades of observations that can be publicly searchable after the initial proprietary period. KOA provides a convenient interface to search these valuable datasets, especially with PyKOA, to batch query and access these data. With the archival information, observers can better make the judgement if an alert is of interest, and if so, can trigger Keck ToO observations with the TORT and take proper calibration data for real-time data reduction for the instrument used to carry out the observations. 

These examples demonstrate the efficacy and efficiency of Keck, both in triggering follow-up and providing data products in a timely manner in aiding further decision making to the time-domain follow-up community.

\section{Summary}
Large telescopes with robust ToO trigger and follow-up capabilities play an important role in the time-domain and multi-messenger astronomy ecosystem, in particular to catch fast evolving events before they fade away. At Keck, we are actively implementing software infrastructure to enable ToO triggers, data reduction, and data sharing in a timely fashion. On the follow-up front, our Target-of-Opportunity Observation Request API will enable automatic ToO triggering and interruption. With the TORT API, Keck can interface with science platforms, such as the TOMtoolkit and SkyPortal, that allow world-wide collaborators to share their target information, observation resources, and data visualization to enable coordinated follow-ups and decision making. 

Keck observations can be easily accessible via the Keck Observatory Archive. KOA provides various methods and services that allow users to query the entire database via the KOA landing page. Users can stream data in real-time as data is acquired with the Observers' Data Access Portal. For large or complex queries, the PyKOA client enables a programmatic approach to search both public and private data. With TORT and KOA, Keck is well-connected to receiving ToO triggers and providing observations and related data products in real-time. Keck is well-poised to be an integral part of the astronomical community to assemble an end-to-end follow-up infrastructure and maximize the science outcome from millions of alerts delivered by ongoing and future TDA/MMA surveys.

\acknowledgments 
W. M. Keck Observatory is operated as a scientific partnership among the California Institute of Technology, the University of California, and the National Aeronautics and Space Administration. The Observatory was made possible by the generous financial support of the W. M. Keck Foundation.

The Keck Observatory Archive is operated by the W. M. Keck Observatory and the NASA Exoplanet Science Institute, under contract with the National Aeronautics and Space Administration.

The authors wish to recognize and acknowledge the very significant cultural role and reverence that the summit of Maunakea has always had within the indigenous Hawaiian community. We are most fortunate to have the opportunity to conduct observations from this mountain.



\end{document}